\documentclass[10pt,conference]{IEEEtran}

\usepackage{cite}
\ifCLASSINFOpdf
  \usepackage[pdftex]{graphicx}
  \graphicspath{{../figures/}}
  \DeclareGraphicsExtensions{.pdf,.jpeg,.png}
\fi

\usepackage{courier}

\usepackage{array}
\usepackage{stfloats}
\usepackage{url}

\begin{document}

\title{Towards Automatic Generation of \\ Short Summaries of Commits}
\author{\IEEEauthorblockN{Siyuan Jiang and Collin McMillan}
\IEEEauthorblockA{Department of Computer Science and Engineering\\
University of Notre Dame\\
Notre Dame, IN, USA\\
Email: \{sjiang1, cmc\}@nd.edu}}
\maketitle

\begin{abstract}
Committing to a version control system means submitting a software change to the system. Each commit can have a message to describe the submission. Several approaches have been proposed to automatically generate the content of such messages. However, the quality of the automatically generated messages falls far short of what humans write. In studying the differences between auto-generated and human-written messages, we found that 82\% of the human-written messages have only one sentence, while the automatically generated messages often have multiple lines. Furthermore, we found that the commit messages often begin with a verb followed by an direct object. This finding inspired us to use a ``verb+object'' format in this paper to generate short commit summaries. We split the approach into two parts: verb generation and object generation. As our first try, we trained a classifier to classify a diff to a verb. We are seeking feedback from the community before we continue to work on generating direct objects for the commits.
\end{abstract}

\section{Introduction}
\label{sec:intro}
A commit is the action of software developers submitting a software change to a version control system. Commits can have commit messages, which are often written by developers to describe the changes. Commit messages are important because developers use them to review, validate, and understand the commits, but commit messages sometimes are non-informative or even empty~\cite{Vasquez2015ICSE}.

To address this problem, automatic commit message generation techniques have been proposed. They often use program analysis and differencing techniques to generate summaries of changes~\cite{Vasquez2015ICSE,Buse2010ASE,Moreno2014FSE,Linares2015ICSE}. These summaries are much shorter than the diff files (generated by differencing tools), but the summaries still tend to have multiple lines. Other techniques generate commit messages from other project documents. For example, Rastkar and Murphy proposed to generate the commit messages from user stories~\cite{Rastkar2013ICSE}.  The summaries generated by these techniques are useful, but what is still missing is one-sentence summaries which convey the key ideas of commits. 

\begin{figure}[!t]
\centering
\includegraphics[width=1.7in]{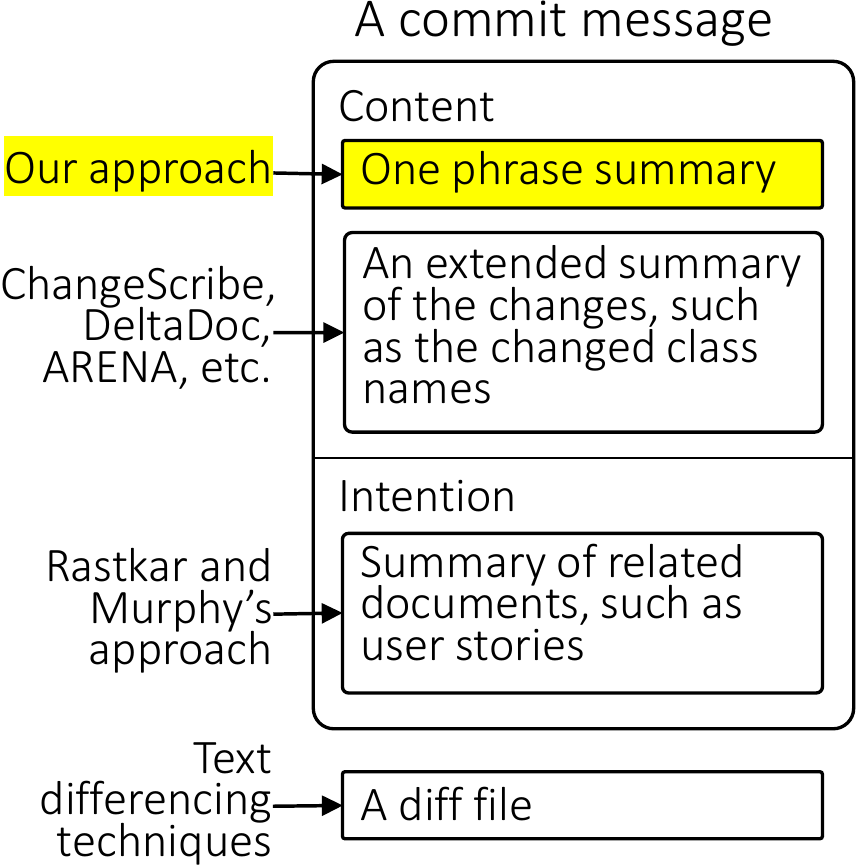}
\vspace{-0.1cm}
\caption{Existing techniques for commit message generation compared to our approach}
\vspace{-0.4cm}
\label{fig:overview}
\end{figure}

The idea of generating one-sentence summaries is based on our exploratory data analysis on the two million commit messages that we present in this paper. We used natural language processing (NLP) techniques to analyze the text of the commit messages and found that the majority of the commit messages are only one sentence long, and \textbf{nearly half of the commit messages begin with a verb followed by a direct object}. This finding inspired us to design a method for generating commit summaries that are similar to what developers write: ``verb + object''. These one-phrase summaries can be the leading sentences or the topics of the summaries generated by the existing techniques (Figure~\ref{fig:overview}).

We divided our work into three parts: the exploratory data analysis, the verb generation and the direct-object generation. We summarized the exploratory data analysis in the previous paragraph and describe it in Section~\ref{sec:exploratory}. For the verb generation, we trained a Naive Bayes classifier to identify verbs based on diff files that should be the key verbs in the commit messages. We also conducted a preliminary evaluation of the verb generation in Section~\ref{sec:verbgen}. 

In this ERA paper, we present several open questions to the community that we hope will guide our future work, and in particular the generation of direct objects for the verbs.

Our contributions include:
\begin{itemize}
  \item Using NLP techniques to analyze the commit messages, which enable us to analyze a large set of the messages (which we release in our online appendix)
  \item Discovery of a common phrase structure that is used by software developers to write commit messages, and a program that automatically extracts such phrase structure
  \item A proposal that aims to generate one-sentence commit messages that convey the key ideas of commits
\end{itemize}

In the rest of this paper, we will present a motivational example, the related work, the exploratory data analysis, the verb generation technique, and the future work.

\emph{Online Appendix}
\label{sec:appendix}
We put our scripts and results on our online appendix: {\small{\textbf{\url{http://nd.edu/~sjiang1/commitact}}}}

\section{Example}
\label{sec:example}
\begin{figure}[!t]
\centering
\includegraphics[width=3.5in]{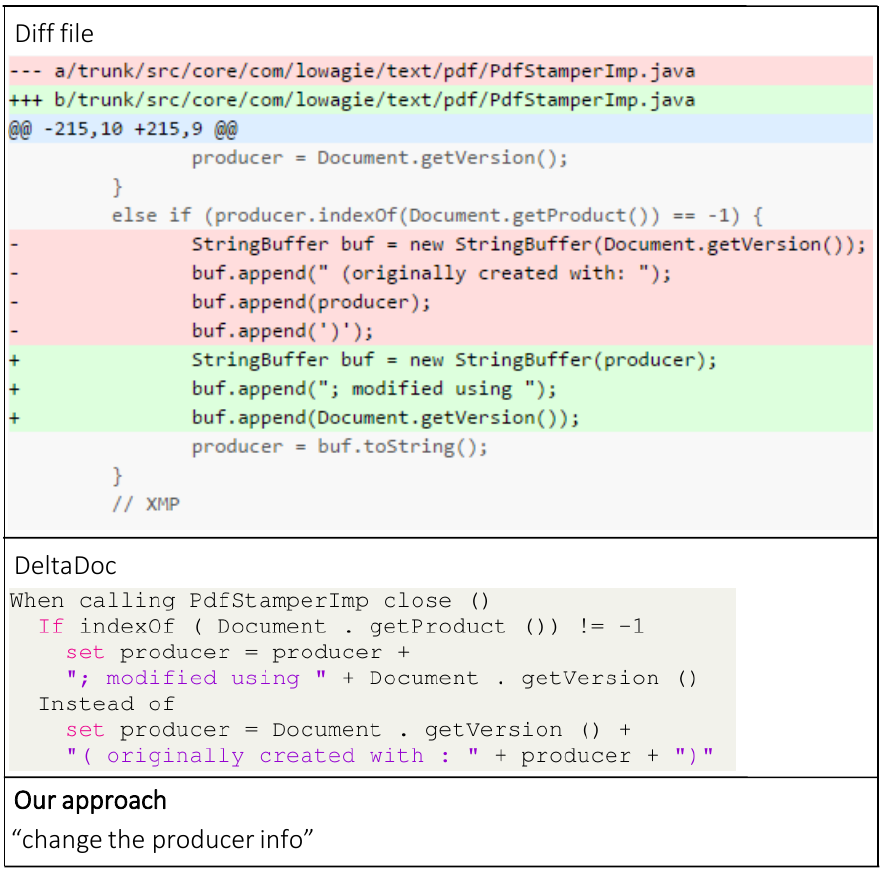}
\caption{The diff file, the commit message generated by DeltaDoc~\cite{Buse2010ASE}, and the commit message that our approach aims to generate for Commit r3909 in iText (Section~\ref{sec:example})}
\label{fig:r3909-diff}
\end{figure}

In this section, we borrow the example of Commit r3909 in iText from the paper of DeltaDoc~\cite{Buse2010ASE}. The diff file of r3909 and the summary generated by DeltaDoc are shown in Figure~\ref{fig:r3909-diff}.

The size of the generated document is about half of the diff file, but it is still difficult to get the general idea at the first glance. Similarly, Changescribe~\cite{Linares2015ICSE} also generates messages that are several lines long. What is missing is a leading sentence that summarizes all the changes in a commit. Now consider the commit message that the developer wrote: ``Changing the producer info.'' This phrase contains the action of the commit, ``change'', and what is the object of the action, ``the producer info''. The developer can skim this phrase and understand what was changed in the commit.

Currently, our approach generates ``change'' for r3909, and in the future, we will have an approach to generate ``the producer info''. The combination of the two approaches is going to generate phrases like ``change the producer info''.

\section{Related Work}
Our project has two parts: exploratory data analysis and commit message generation. Based on the two parts, we separate the related work into three categories: empirical studies about commit messages, empirical studies about diff files, and techniques that generate commit messages.

\subsection{Empirical Studies about Commit Messages}
Several empirical studies about commits messages have been conducted for commit classification and commit message generation~\cite{Moreno2014FSE,Buse2010ASE,Alali2008ICPC,Hattori2008ASE,Hindle2009ICPC}. 
For example, Moreno et al.~\cite{Moreno2014FSE} manually inspected the existing release notes before they designed an approach to generate release notes automatically. Buse and Weimer~\cite{Buse2010ASE} conducted a similar manual inspection for automatic commit message generation. Like these previous studies, our exploratory data analysis aims to gain insights for our approach of generating commit messages.

Different from the previous studies, we used natural language processing (NLP) techniques, which help us to mine information from the existing commit messages automatically and confirm hypotheses on a large data set. Besides manual inspection, the previous studies also computed the sizes of commit messages and analyzed the messages as bags of words~\cite{Hattori2008ASE,Hindle2009ICPC}. 
In contrast, we are able to conduct grammar analysis on the commit messages. The grammar analysis leaded to a key finding that shaped our approach.

\subsection{Empirical Studies about Commit Changes}
There are many empirical studies about the changes in commits~\cite{Fluri2007TSE,Hindle2009ICPC,Alali2008ICPC,Hattori2008ASE}. For example, Fluri et al. studied change types based on their syntax differencing technique~\cite{Fluri2007TSE}. Currently, we have not conducted an empirical study on the commit changes, but we plan to study the content of the diff files in the future. Instead of looking for change types, we will study whether there are overlapped words in the commit messages and their diff files and where we can locate the overlapped words in the diff files.

\subsection{Commit Message Generation Techniques}
A common way to generate commit messages is summarizing code changes of a commit~\cite{Linares2015ICSE,Moreno2014FSE,Buse2010ASE}. Many techniques use syntax differences to present code changes~\cite{Fluri2007TSE,Linares2015ICSE,Moreno2014FSE,Buse2010ASE}. Different from the existing techniques, we use diff files (generated by git diff command) in our approach. Diff files are textual differences and easy to obtain. On the other hand, syntax differencing requires code parsing. Additionally, syntax differencing includes only code changes, while diff files contain other changes, such as comment and makefile changes. While the two differencing types have their own advantages and disadvantages, we chose to use the diff files as our first try, because they are easier to obtain.

To include context information in a commit message, several approaches consider the information outside the text or code changes of a commit~\cite{Moreno2014FSE,Le2015ICPC,Rastkar2013ICSE}. While we agree that the context information is an important part of a commit message, our approach is currently focusing on summarizing text changes into a short sentence to increase readability and interpretability of a commit message.

Our approach to generate a verb for a commit is similar to the approach taken by Le et al. to link issue reports to commits~\cite{Le2015ICPC}. Le et al. conducted textual similarity analysis between commit messages and issue reports where they used term frequency-inverse document frequency (tf-idf) to represent commit messages and issue reports. We also used tf-idf, but tf-idf is used to represent the diff files instead of the commit messages.

\section{Exploratory Data Analysis}
\label{sec:exploratory}
We conducted an exploratory data analysis that is similar to the analysis done by Hattori and Lanza~\cite{Hattori2008ASE}. Hattori and Lanza found that most commits include few files and very few commits have hundreds of files. Likewise, we found that most commit messages have few sentences and few commit messages have more than ten sentences.

\emph{The Data Set}
First, we obtained 967 commits from the work by Mauczka et al~\cite{Mauczka2015MSR}. Second, we obtained all the commits from the top 1,000 popular Java projects in Github (due to space limit, we put the details on our online appendix, Section~\ref{sec:appendix}). Then, we filtered the commit messages that are empty or have non-English letters. In the end, we obtained 2,027,734 commits.

\emph{Removing Special Commits}
We excluded the rollback and merge commits from our analysis. Version control systems often provide automatic commit messages for rollbacks and merges, such as, ``merge commits X and Y''. In the two million commits, we removed nearly 400k rollbacks and merges by checking whether the commit messages are begun with ``merge'' or ``rollback''. 

\begin{figure}[!t]
\centering
\includegraphics[width=2.2in]{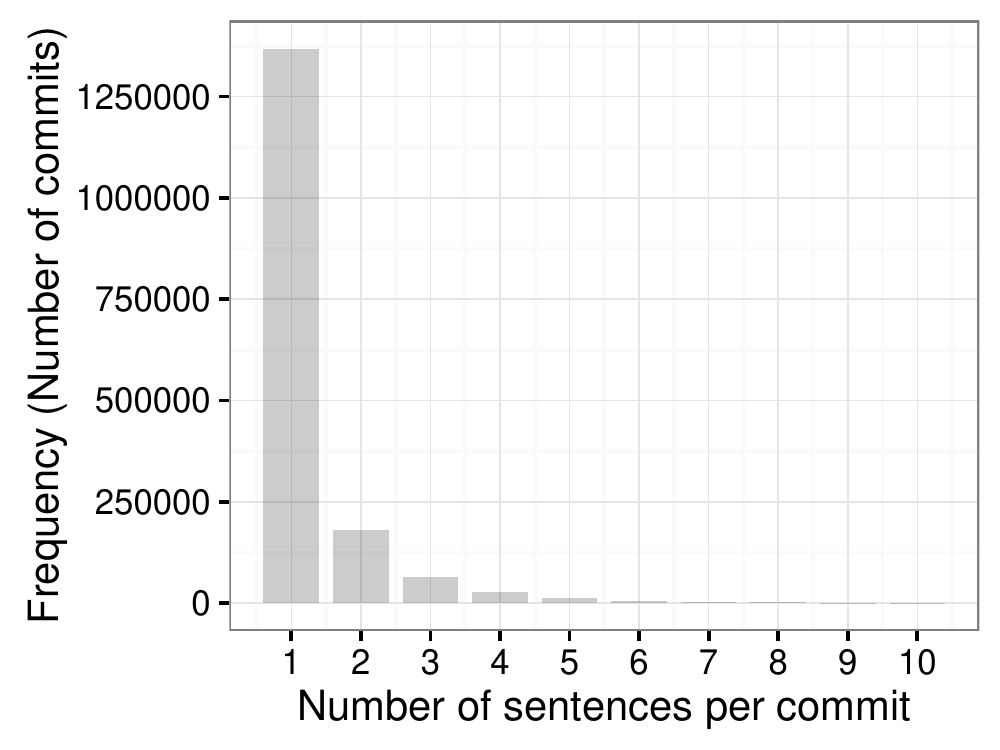}
\caption{Histogram of number of sentences in the commit messages}
\label{fig:hist-numSentences}
\end{figure}

\emph{Number of the Sentences}
In the remaining 1.6 million commit messages, we counted the number of the sentences in each commit message by using Stanford CoreNLP~\cite{Manning2014ACL}. The majority of the commit messages have few sentences. 82\% of the commit messages have only one sentence. Only 0.2\% of the commit messages have more than ten sentences. Figure~\ref{fig:hist-numSentences} shows the histogram of the number of the sentences in the commit messages (excluding the messages have more than ten sentences due to space limit). 

\emph{Grammar Analysis on the Commit Messages}
\label{subsec:grammaranalysis}
We took two steps in the grammar analysis. First, we manually read 12 randomly-sampled commit messages from the commits we obtained from Mauczka et al~\cite{Mauczka2015MSR}. In this step, we formed the hypothesis that ``verb + object'' is a common phrase structure in the commit messages. Second,  to confirm the hypothesis, we used Stanford CoreNLP~\cite{Manning2014ACL} to detect the verbs and their direct objects in the first sentences of the commit messages. In the 1.6 million messages, we found 763,826 messages (which is 47\% of the 1.6 million messages) where the first sentences are begun with a verb and its direct object.

\section{Classifying Diffs into Verb Groups}
\label{sec:verbgen}
In this section, our goal is to generate a verb from a commit. We used diff files (i.e., textual differences) to represent the changes of commits because diff files can be easily obtained by git diff command. Then we treated the problem of verb generation as a multiclass classification problem---classifying a diff file into one of the verb groups, where a verb group is a group of verbs that have similar meanings. As the first step, we define our verb groups in the following section. 

\subsection{Verb Groups}
\begin{figure}[!t]
\centering
\includegraphics[width=2.4in]{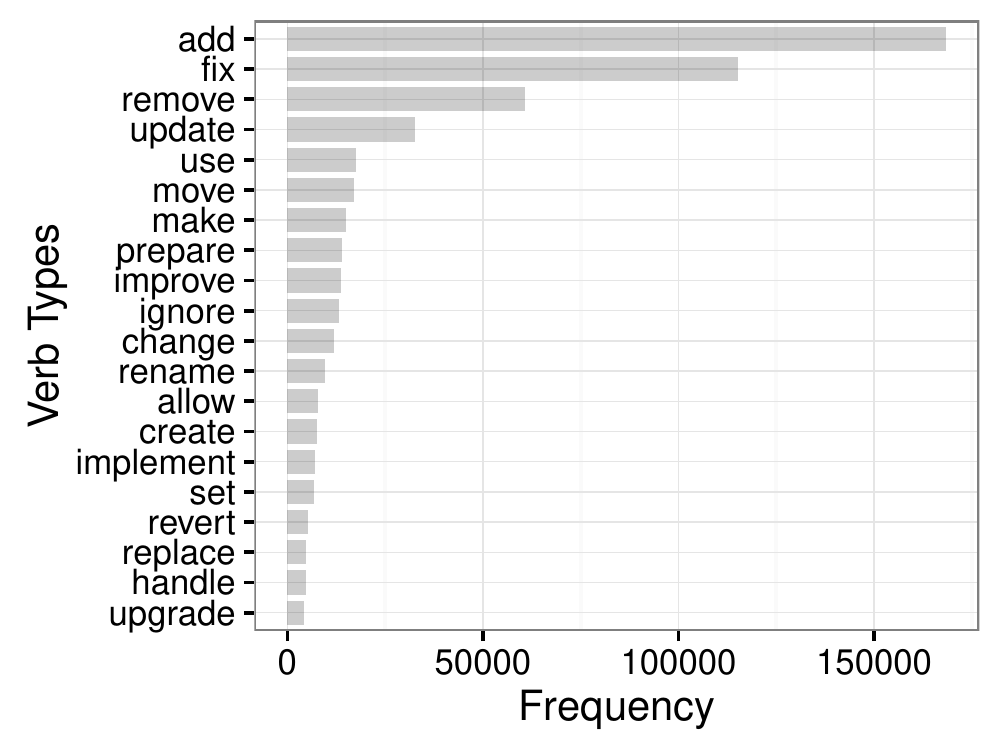}
\caption{Histogram of the verbs in commit messages}
\label{fig:hist-verbs}
\end{figure}

\vspace{-0.1cm}
When we analyzed phrase structures of the commit messages (Section~\ref{subsec:grammaranalysis}), we retrieved for each commit a verb from the commit message. There are 763k verbs in total. We transformed the verbs into their lemmas and we called each distinct lemma a verb type. There are 4962 verb types in the 763k verbs. Figure~\ref{fig:hist-verbs} shows the histogram of the 20 most frequent verb types. Alali et al.~\cite{Alali2008ICPC} has reported a list of frequent words in commit messages, which overlap with our frequent verb types.

From all the verb types, we considered only the 20 most frequent word types, which cover 70\% of the commit messages (537k commit messages). We grouped similar word types by using a word embedding tool\footnote{\url{http://bionlp-www.utu.fi/wv_demo/} On this webpage, we looked up the 20 nearest words for each verb type. Two verb types are grouped together if one verb type is in the other verb type's 20 nearest word list.}, which uses word2vec method~\cite{Mikolov2013ANIP}. Finally we manually inspected the grouped verbs and added ``implement'' to the group of ``add''. There are 15 verb groups in total, which are shown in Table~\ref{tab:vgroups}. The first, third, and fifth columns list the ids of the verb groups.

\begin{table}[!t]
\caption{The Verb Groups}
\vspace{-0.2cm}
\label{tab:vgroups}
\centering
\begin{tabular}{llllll}
\hline
Id & Verb types & Id & Verb types & Id & Verb types \\
\hline
1 & add, create, make,  &  6 & move, change &  12 & allow\\
   & implement              &  7 & prepare           & 13 & set \\
2 & fix                           &  8 & improve           & 14 & revert \\
3 & remove                 & 9 & ignore                &  15 & replace \\
4 & update, upgrade   & 10 & handle             & &  \\
5 & use                        & 11 & rename           &  & \\
\hline
\end{tabular}
\vspace{-0.3cm}
\end{table}

\emph{Labeling}
\label{labelling}
To label each diff file, we used the verb that we extracted from the commit message, and we labeled the diff file with the verb group id that includes the verb. The verb groups only include the 20 most frequent verb types, in this study, we excluded the diffs that have other verbs. In total, we have 537k labeled diff files.

\subsection{The Data Set}
We removed the diff files that are larger than 1MB due to space limit. We also removed the diff files that have non-ascii codes. In the end, we have 509k labeled diff files. We randomly selected 3k diff files as the test set and the rest of the diff files are used for training.

\subsection{Overall Approach}
\begin{figure}[!t]
\centering
\includegraphics[width=2.3in]{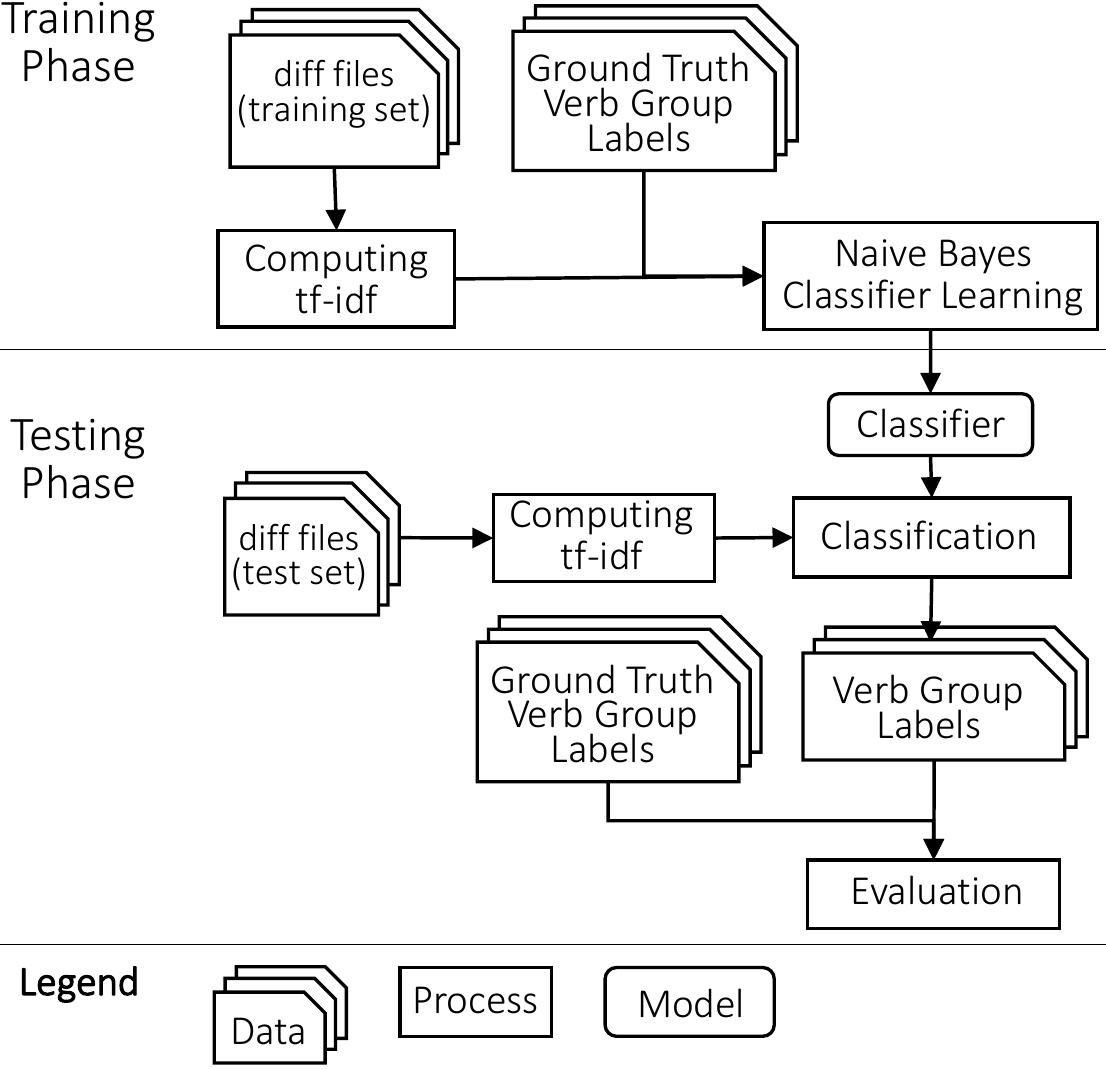}
\caption{The overall approach}
\vspace{-0.3cm}
\label{fig:approach}
\end{figure}

The overall approach is shown in Figure~\ref{fig:approach}. We chose a Naive Bayes classifier to classify the diff files into the verb groups. Before we train the classifier on the diff files, we computed tf-idf (term frequency-inverse document frequency) for every word type (i.e., distinct word) as the features of the diff files. Tf-idf is a common textual feature that evaluates the importance of a word type by two factors: 1) the number of times the word type occurs in a diff file divided by the total number of words in the diff file, and 2) the number of times the word type occurs in all the diff files~\cite{Le2015ICPC}. 

\subsection{Evaluation}
The overall accuracy is 39\%; the precision is 43\%; and the recall is 39\%. The classifier works best for verb groups 1 and 9. The precision for verb group 1 is 38\% and the recall is 100\%; the precision for verb group 9 is 100\% and the recall is 41\%. Although we trained the classifier with 15 verb groups, the classifier classified the test set into five verb groups and was not able to detect any of the other ten verb groups. We plan to improve our training approach by 1) trying other machine learning techniques, such as random forests~\cite{Le2015ICPC}; 2) using SMOTE~\cite{smote} to address the problem of the unbalanced data set (most of the diffs are labeled with verb group 1).

\section{Discussion and Future Work}
In the process of this project, we have formed several potential research questions to be discussed in the conference. We hope the conversions at the conference will help in directing us towards answering these questions.

\emph{RQ1} What techniques are appropriate for generating direct objects for the commits?
We observed that the direct objects often occur in the diff files. So one of our options is to use extractive summarization techniques to extract the ``direct objects'' from the diff files.

\emph{RQ2} What machine learning models and features suit verb-generation task better? 
To improve our verb-generation approach, we can try other classification methods, such as decision trees. 
Feature-wise, diff files follow a certain format and we can create some features to represent the characteristics of a diff file, for example, the number of ``+'' in a diff file. 

\emph{RQ3} To what extent are the short summaries useful?
Although we think the short summaries are useful based on our experience, we need to conduct a study to confirm our hypothesis. Our current assumption is that the short summaries help developers understand a commit more quickly. 


\section*{Acknowledgment}
This work was partially supported by the NSF CCF-1452959 and CNS-1510329 grants, and the Office of Naval Research grant N000141410037. Any opinions, findings, and conclusions expressed herein are the authors' and do not necessarily reflect those of the sponsors.

\bibliographystyle{IEEEtran}
\bibliography{main}

\begin{thebibliography}{10}
\providecommand{\url}[1]{#1}
\csname url@samestyle\endcsname
\providecommand{\newblock}{\relax}
\providecommand{\bibinfo}[2]{#2}
\providecommand{\BIBentrySTDinterwordspacing}{\spaceskip=0pt\relax}
\providecommand{\BIBentryALTinterwordstretchfactor}{4}
\providecommand{\BIBentryALTinterwordspacing}{\spaceskip=\fontdimen2\font plus
\BIBentryALTinterwordstretchfactor\fontdimen3\font minus
  \fontdimen4\font\relax}
\providecommand{\BIBforeignlanguage}[2]{{%
\expandafter\ifx\csname l@#1\endcsname\relax
\typeout{** WARNING: IEEEtran.bst: No hyphenation pattern has been}%
\typeout{** loaded for the language `#1'. Using the pattern for}%
\typeout{** the default language instead.}%
\else
\language=\csname l@#1\endcsname
\fi
#2}}
\providecommand{\BIBdecl}{\relax}
\BIBdecl

\bibitem{Vasquez2015ICSE}
M.~Linares-V\'{a}squez, L.~F. Cort\'{e}s-Coy, J.~Aponte, and D.~Poshyvanyk,
  ``Changescribe: A tool for automatically generating commit messages,'' in
  \emph{ICSE '15}, vol.~2, May 2015, pp. 709--712.

\bibitem{Buse2010ASE}
R.~P. Buse and W.~R. Weimer, ``Automatically documenting program changes,'' in
  \emph{ASE '10}, 2010, pp. 33--42.

\bibitem{Moreno2014FSE}
L.~Moreno, G.~Bavota, M.~Di~Penta, R.~Oliveto, A.~Marcus, and G.~Canfora,
  ``Automatic generation of release notes,'' in \emph{Proceedings of the 2014
  FSE}, pp. 484--495.

\bibitem{Linares2015ICSE}
M.~Linares-V\'{a}squez, L.~F. Cort\'{e}s-Coy, J.~Aponte, and D.~Poshyvanyk,
  ``Changescribe: A tool for automatically generating commit messages,'' in
  \emph{2015 IEEE/ACM 37th IEEE ICSE}, vol.~2, pp. 709--712.

\bibitem{Rastkar2013ICSE}
S.~Rastkar and G.~C. Murphy, ``Why did this code change?'' in \emph{Proceedings
  of the 2013 ICSE}, ser. ICSE '13, 2013, pp. 1193--1196.

\bibitem{Alali2008ICPC}
A.~Alali, H.~Kagdi, and J.~I. Maletic, ``What's a typical commit? a
  characterization of open source software repositories,'' in \emph{2008 16th
  IEEE Intl. Conf. on Program Comprehension}, pp. 182--191.

\bibitem{Hattori2008ASE}
L.~P. Hattori and M.~Lanza, ``On the nature of commits,'' in \emph{2008 23rd
  ASE - Workshops}, pp. 63--71.

\bibitem{Hindle2009ICPC}
A.~Hindle, D.~M. German, M.~W. Godfrey, and R.~C. Holt, ``Automatic
  classication of large changes into maintenance categories,'' in \emph{2009
  IEEE 17th ICPC}, pp. 30--39.

\bibitem{Fluri2007TSE}
B.~Fluri, M.~Wuersch, M.~PInzger, and H.~Gall, ``Change distilling:tree
  differencing for fine-grained source code change extraction,'' \emph{IEEE
  TSE}, vol.~33, no.~11, pp. 725--743, 2007.

\bibitem{Le2015ICPC}
T.~D.~B. Le, M.~Linares-Vasquez, D.~Lo, and D.~Poshyvanyk, ``Rclinker:
  Automated linking of issue reports and commits leveraging rich contextual
  information,'' in \emph{2015 IEEE 23rd ICPC}, pp. 36--47.

\bibitem{Mauczka2015MSR}
A.~Mauczka, F.~Brosch, C.~Schanes, and T.~Grechenig, ``Dataset of
  developer-labeled commit messages,'' ser. MSR '15, pp. 490--493.

\bibitem{Manning2014ACL}
C.~D. Manning, M.~Surdeanu, J.~Bauer, J.~Finkel, S.~J. Bethard, and
  D.~McClosky, ``The {Stanford} {CoreNLP} natural language processing
  toolkit,'' in \emph{ACL System Demonstrations}, 2014, pp. 55--60.

\bibitem{Mikolov2013ANIP}
T.~Mikolov, I.~Sutskever, K.~Chen, G.~S. Corrado, and J.~Dean, ``Distributed
  representations of words and phrases and their compositionality,'' in
  \emph{NIPS}, 2013, pp. 3111--3119.

\bibitem{smote}
N.~V. Chawla, K.~W. Bowyer, L.~O. Hall, and W.~P. Kegelmeyer, ``Smote:
  synthetic minority over-sampling technique,'' \emph{Journal of artificial
  intelligence research}, vol.~16, pp. 321--357, 2002.

\end{thebibliography}
\end{document}